\documentclass{article}

\usepackage{PRIMEarxiv}
\usepackage{amsmath}
\usepackage[utf8]{inputenc} 
\usepackage[T1]{fontenc}    
\usepackage{hyperref}       
\usepackage{url}            
\usepackage{booktabs}       
\usepackage{amsfonts}       
\usepackage{nicefrac}       
\usepackage{microtype}      
\usepackage{lipsum}
\usepackage{fancyhdr}       
\usepackage{graphicx}       
\graphicspath{{media/}}     

\title{Tone recognition in low-resource languages of North-East India: peeling the layers of SSL-based speech models
\thanks{\textit{\underline{Citation}}: 
\textbf{Authors. Title. Pages.... DOI:000000/11111.}} 
}
\author{
  {\normalfont Parismita Gogoi$^{1,2}$, 
  Sishir Kalita$^{3}$, 
  Wendy Lalhminghlui$^{1}$, 
  Viyazonuo Terhiija$^{1}$,} \\
  {\normalfont Moakala Tzudir$^{4}$, 
  Priyankoo Sarmah$^{1}$, 
  S. R. M. Prasanna$^{5}$} \\[0.5em]
  \footnotesize
  $^{1}$IIT Guwahati, India \\
  $^{2}$DUIET, Dibrugarh University, India \\
  $^{3}$Armsoftech.air, India \\
  $^{4}$National Institute of Electronics \& Information Technology, Kohima, India \\
  $^{5}$IIIT Dharwad, India \\[0.3em]
  \texttt{parismitagogoi@iitg.ac.in, priyankoo@iitg.ac.in}
}

\begin{document}
\maketitle

\begin{abstract}
This study explores the use of self-supervised learning (SSL) models for tone recognition in three low-resource languages from North Eastern India: Angami, Ao, and Mizo. We evaluate four Wav2vec2.0 base models that were pre-trained on both tonal and non-tonal languages. We analyze tone-wise performance across the layers for all three languages and compare the different models. Our results show that tone recognition works best for Mizo and worst for Angami. The middle layers of the SSL models are the most important for tone recognition, regardless of the pre-training language, i.e. tonal or non-tonal. We have also found that the tone inventory, tone types, and dialectal variations affect tone recognition. These findings provide useful insights into the strengths and weaknesses of SSL-based embeddings for tonal languages and highlight the potential for improving tone recognition in low-resource settings.
\end{abstract}
\section{Introduction}
This paper reports an attempt at tone recognition in three low-resource languages of North-east India, namely, Angami, Ao, and Mizo, leveraging pre-trained self-supervised learning (SSL)-based models created in high-resource languages. This work has two major outcomes; first, it demonstrates that pre-trained SSL models achieve considerable accuracy in tone recognition in the low-resource languages investigated in this study. Secondly, the current work reveals that tone recognition is more effective in the middle layers, specifically between the 4th and the 6th layers of the network, which makes us believe that these layers may exclusively model tone-specific information. As a tertiary finding, we also observed that tone inventory size and tone types also have an effect on the accuracy of tone recognition when using pre-trained models. 


The languages in the current study, namely, Angami, Ao, and Mizo, are low-resource lexical tone languages without any large labelled corpora. At the same time, accuracy in tone recognition is crucial for developing language technology in lexical tone languages. Considering that, the current study focuses on leveraging SSL-based representations to enhance the accuracy of tone recognition performance in these languages. Several Tibeto-Burman languages have already attempted to develop tone recognition systems, including languages such as Ao and Mizo~\cite{gogoi-EtAl:2020:LREC, gogoi2020automaticAo, gogoi2021learning}. However, these works have relied on hand-crafted acoustic features derived primarily from fundamental frequency ($f_{0}$) contours of tones~\cite{ gogoi-EtAl:2020:LREC, gogoi2020automaticAo, gogoi2021learning, biswa15, devi2024disambiguation}. While such approaches have yielded considerable accuracy in tone identification, they depend entirely on manual annotation of speech data with careful tones marking. 


The existing works on automatic tone recognition primarily focus on Mandarin Chinese, using a range of acoustic front-ends, such as Mel Frequency Cepstral Coefficients (MFCCs), Mel-spectrograms, $f_{0}$, energy, etc. Classical methods including Gaussian Mixture Model (GMM), Support Vector Machines (SVMs), and Hidden Markov Models (HMMs) have successfully performed tone recognition on segmented syllables in Mandarin Chinese \cite{chen1987hidden, yang1988hidden, liu1989tone, lee2002using, peng2005tone}. Neural networks have further advanced tone modeling in Mandarin Chinese, with Multi-Layer Perceptron (MLP)-based recognition of monosyllables, using $f_{0}$ and energy features \cite{chang1990mandarin, Ryant14highlyaccurate, Ryant_mandarintone}. Recently, Convolutional Neural Network (CNN), Recurrent Neural Network (RNN), and further transformer-based approaches were explored in Mandarin tone recognition handcrafted features \cite{Chen+2016, gao2019tonenet, tang2021end, huang2021encoder, liu2024learning, liu2023jtone}. While these approaches are successful, they are dependent on the large amount of speech data in Mandarin Chinese.

SSL models have shown that the dependence on manually annotated speech data can be reduced in modeling suprasegmental features. SSL models such as Wav2Vec2.0~\cite{baevski2020wav2vec} and HuBERT~\cite{hsu2021hubert} have demonstrated their capabilities in successfully encoding suprasegmental features such as tone, stress, accent, and intonation~\cite{de2024layer, yang2023can}. These models utilize large-scale unlabeled speech data to learn representations which is beneficial for languages with limited resources~\cite{mohamed2022self}. The application of such models is demonstrated in various low-resource languages~\cite{zhao2022improving, mdhaffar2024performance}.

Some of the recent studies have provided insight into how tone is represented in the hidden layers of self-supervised models~\cite{de2024layer, shen2024encoding}. It was shown that models pre-trained in Mandarin, Vietnamese, or even in non-tonal languages, such as English, can capture tones effectively in the middle layers. While fine-tuning such models with tone languages improves the performance of tone recognition, fine-tuning with non-tone languages does not yield any improvement in tone recognition~\cite{de2024layer, shen2024encoding}. This motivated the current work to use SSL models trained in both tonal and non-tonal languages to build tone recognition systems in Angami, Ao, and Mizo languages and to explore the layers that provide better tone recognition. 



The emergence of SSL-based representations, trained on diverse large-scale datasets, provides an opportunity to learn the rich hierarchical features directly from raw waveforms~\cite{mohamed2022self}. In the current study, considering the diverse tone types and tone inventory size of Angami, Ao, and Mizo, we explore how the models perform in recognizing the tones. We systematically evaluate the efficacy of the Wav2Vec2.0 model trained in Mandarin Chinese, Vietnamese, and English databases, built on various sizes of speech data. Finally, the current work provides:
\begin{itemize}
    \item Layer-wise analysis in the tone recognition task
    \item Speaker-independent accuracy in tone recognition
    \item Cross-dialectal performance in tone recognition.
\end{itemize}





\section{Methodology}
This section outlines the methodology of the paper. It begins by describing the three low-resource languages used in the study, followed by an overview of the developed speech database. Next, the pre-trained models are briefly explained, and finally, the classification model along with its performance evaluation is presented.

\subsection{Languages of the current study}
The three languages in the current study belong to the Tibeto-Burman family of languages. The Tibeto-Burman family of languages usually have a varying number and types of lexical tones. 

Angami, also known as Tenyidie, is a low-resource language spoken in the state of Nagaland in India. Angami has four distinct level tones that are represented by numbers. The tones in Angami are \emph{T1}, \emph{T2}, \emph{T3} and \emph{T4}, in descending order of $f_0$ height, which correspond to high, mid, mid-low, and low, respectively~\cite{terhiija2024voice}. Ao, another Tibeto-Burman language, is spoken in Nagaland, India. Ao has three three primary dialects, namely, Changki, Chungli, and Mongsen. Each dialect assigns one of the three tones, High (\emph{H}), Mid (\emph{M}), and Low (\emph{L}) to the Ao syllables~\cite{tzudir2021analysis, tzudir2023thesis}. Mizo is spoken in the Mizoram state of India and in parts of Bangladesh and Myanmar. Mizo has four lexical tones, High (\emph{H}), Low (\emph{L}), Rising (\emph{R}), and Falling (\emph{F}) \cite{sarmah2010preliminary, wendy2023thesis}. Among the three languages in the current study, only Mizo has tones with dynamic $f_0$ contours.




\begin{table*}[!t]
\centering
\caption{Distribution of tone tokens for Angami, Ao, and Mizo after computing the embeddings using the 0.05 sec duration threshold.}
\label{table:dataset}
\resizebox{\textwidth}{!}{
\begin{tabular}{|c|ccccc|cccc|ccccc|}
\hline
\multicolumn{1}{|l|}{} & \multicolumn{5}{c|}{Angami}                                                                                                       & \multicolumn{4}{c|}{Ao}                                                                             & \multicolumn{5}{c|}{Mizo}                                                                                                      \\ \hline
Tone                   & \multicolumn{1}{c|}{T1}    & \multicolumn{1}{c|}{T2}    & \multicolumn{1}{c|}{T3}    & \multicolumn{1}{c|}{T4}   & Total          & \multicolumn{1}{c|}{L}    & \multicolumn{1}{c|}{M}     & \multicolumn{1}{c|}{H}    & Total          & \multicolumn{1}{c|}{L}    & \multicolumn{1}{c|}{H}    & \multicolumn{1}{c|}{R}    & \multicolumn{1}{c|}{F}    & Total          \\ \hline
\#Tokens               & \multicolumn{1}{c|}{3699}  & \multicolumn{1}{c|}{6323}  & \multicolumn{1}{c|}{5404}  & \multicolumn{1}{c|}{3369} & \textbf{18795} & \multicolumn{1}{c|}{5426} & \multicolumn{1}{c|}{7367}  & \multicolumn{1}{c|}{4585} & \textbf{17378} & \multicolumn{1}{c|}{3040} & \multicolumn{1}{c|}{3643} & \multicolumn{1}{c|}{2551} & \multicolumn{1}{c|}{3816} & \textbf{13050} \\ \hline
Duration (mins)        & \multicolumn{1}{c|}{10.17} & \multicolumn{1}{c|}{17.44} & \multicolumn{1}{c|}{12.75} & \multicolumn{1}{c|}{7.73} & \textbf{48.09} & \multicolumn{1}{c|}{8.36} & \multicolumn{1}{c|}{10.55} & \multicolumn{1}{c|}{6.60} & \textbf{25.51} & \multicolumn{1}{c|}{6.29} & \multicolumn{1}{c|}{8.63} & \multicolumn{1}{c|}{6.01} & \multicolumn{1}{c|}{7.75} & \textbf{30.45} \\ \hline
\end{tabular}}
\end{table*}

\subsection{Speech databases}




The Angami speech corpus comprises four dialects: Kohima, Viswema, Kigwema, and Jotsoma village varieties. Data were collected from a total of 69 speakers, of which 30 were female and 39 were male speakers. The target tone is produced in monosyllables and disyllabic words. The native speakers were instructed to read each word in three contexts, namely sentence frames, carrier phrases, and isolation. This study used 20,044 Angami tone tokens. 


The Ao Speech Corpus was developed to study the three dialects which has notable differences in tone assignments. Data was collected from 12 native speakers for each dialect, equally divided between male and female participants for two sessions. Forty trisyllabic words were selected as target words for each dialect, resulting in 2,880 utterances per dialect. These words were read and recorded in three contexts: within sentences, in isolation, and within carrier phrases generating a total of 25,920 tokens.

The speech data for the Mizo Speech Corpus used in the present study is obtained from recordings from 19 Mizo native speakers comprising 10 males and 9 females. Their average is 22 years. The corpus consists of 64 distinct trisyllabic tone combinations of the four Mizo tones. Each combination features five unique phrases, and the medial tone of each phrase is considered in this study. Each phrase is recorded three times per speaker, resulting in 18,240 tone samples.

The speech data for Angami, Ao, and Mizo were recorded using a TASCAM DR-100 MKII digital recorder paired with a Shure SM10A microphone, ensuring high-quality audio at a 44.1 kHz sampling rate. After recording, tone boundaries were meticulously annotated using Praat software \cite{Boersma2020}. In all three languages, tone features were extracted from the Tone-Bearing Units (TBUs), which consisted of the syllable rimes, i.e., vowel nucleus or vowel with a sonorant coda. It is to be noted that for each language, the number of tone samples was reduced for the final experiment as the preprocessing steps, mentioned in Section \ref{sec:pretrained}, forced us to remove some samples. 



\subsection{Pre-trained models explored for analysis}\label{sec:pretrained}
This study focuses exclusively on base Wav2vec2.0 \cite{baevski2020wav2vec} pre-trained models to ensure consistency in architecture and comparable training patterns across evaluations. We selected four Wav2vec2.0 base models, varying in their pre-training datasets, to analyze their effectiveness for tone recognition in Angami, Ao, and Mizo. Our selection is guided by the contrasting characteristics of tonal versus non-tonal monolingual pre-trained models, enabling a comprehensive evaluation of their suitability for tonal language tasks.

\begin{itemize}
    \item \textbf{wav2vec2-base}\footnote{\tiny \url{https://huggingface.co/facebook/wav2vec2-base}}: Pre-trained on 960 hours of Librispeech data, this Wav2Vec2.0-based model is designed for robust speech representation learning from raw audio.
    \item \textbf{mandarin-wav2vec2}\footnote{\tiny\url{https://github.com/kehanlu/mandarin-wav2vec2}}: A Wav2Vec2.0 model pre-trained on 1000 hours of Mandarin audio.
    \item \textbf{chinese-wav2vec2-base}\footnote{\tiny\url{https://huggingface.co/TencentGameMate/chinese-wav2vec2-base}}: This Wav2Vec2.0-based model is pre-trained on the WenetSpeech L subset, comprising 10,000 hours of Mandarin audio.
    \item \textbf{wav2vec2-base-vi}\footnote{\tiny\url{https://huggingface.co/nguyenvulebinh/wav2vec2-base-vi}}: Leveraging 13,000 hours of Vietnamese YouTube audio, this model is pre-trained with the Wav2Vec2.0 base architecture. 
\end{itemize}

To analyze the tone recognition capabilities of SSL-based models, embeddings are extracted from the pre-trained models' layer representations. A minimum duration threshold of 50 milliseconds is applied to tone-labeled segments to ensure meaningful temporal context for feature extraction. After computing the embeddings using the milliseconds duration threshold, the number of tokens for the three languages is shown in Table~\ref{table:dataset}. For each valid tone token, embeddings are extracted across all layers of the models. The embeddings from the corresponding frames within a segment are averaged to produce a single 746-dimensional feature vector per tone token per layer. This dimensionality remains consistent across all models and layers, facilitating direct comparison.


\subsection{Classification model and performance evaluation}
The tone recognition task is framed as a classification problem, with tone labels serving as the target classes. A linear SVM is employed as the classifier. Linear SVM is chosen to simplify the experimental design, focusing the analysis on the ability of SSL embeddings to encode tone-specific information rather than introducing complexity in the classifier itself. This choice ensures that performance differences can be attributed to the quality of the embeddings and not to the sophistication of the classification model. A speaker-independent (SI) Cross-validation (CV) strategy is employed to evaluate the generalizability of the embeddings. The dataset is divided into four folds, each containing tone tokens from distinct speakers, ensuring no overlap between the training and testing sets. This setup tests the embeddings' ability to generalize across unseen speakers. Similar to SI CV, we have also performed a Dialect independent (DI) CV for tone recognition in the case of Ao and Angami, where dialect labels are present. In this case, we considered two dialects' data for training and one dialect's data for testing in each fold.  A stratified cross-validation approach is adopted to balance the class distribution across folds, which ensures that each fold maintains a proportional representation of tone classes. The SVM model is trained on three folds for each iteration and evaluated on the remaining fold. Tone recognition performance is assessed using accuracy and F1-score. For each fold, the accuracy and macro-averaged F1-score are computed. The final performance for each layer and model is reported as the mean and standard deviation of the scores across all folds. This approach enables a layer-wise comparison of tone recognition performance, highlighting the layers that provide the most cues for tone discrimination. 



\section{Experimental results and discussion}
\begin{figure*}[t]
\begin{center}
\includegraphics[scale=0.47]{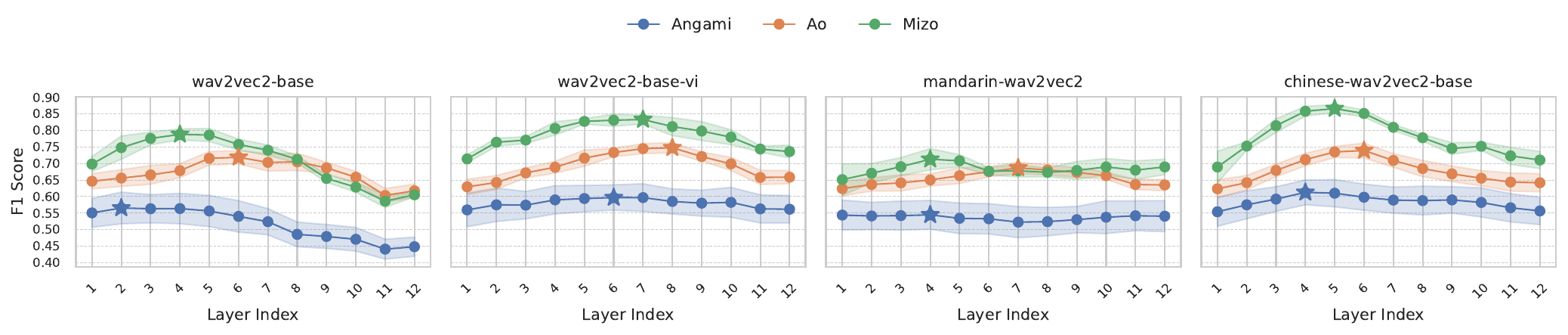}
\caption{Layer-wise comparison of F1-score for three languages (Angami, Ao, Mizo) across various models. The highest scoring layer per language is marked with a star.}
\label{fig:model-wise-perf}
\end{center}
\vspace{-0.3cm}
\end{figure*}

In this section, we comprehensively analyse our experiments on tone recognition in the three tonal languages.
We analyze their performance in different layers to understand how tonal features are represented within the models and to identify which layers are crucial in the tone recognition tasks. First, we study the performance of the tone recognition system developed using the SI CV approach, and later, how dialectal information affects the tone recognition performance is analyzed for Angami and Ao.
Figure~\ref{fig:model-wise-perf} shows the layer-wise performance of various models,  in terms of F1-scores, across the three languages.
 The solid lines represent the mean F1-score in each layer, while the shaded regions depict the standard deviation, indicating the variability in performance. The asterisks highlight the layer with the highest F1-score for each language, indicating the optimal layer for tone classification. 
\subsection{Overall performance in tone recognition }
The results demonstrate significant variations in performance among the models for the languages considered. It is noticed that \textit{chinese-wav2vec2-base} and \textit{wav2vec2-base-vi} based models perform better in the tone recognition task of the current study. As the \textit{chinese-wav2vec2-base} and the \textit{wav2vec2-base-vi} models consist of Mandarin Chinese and Vietnamese tones, it is obvious that tone recognition is better with these two models. However, it is interesting to note that in spite of being trained in a non-tonal language, the \textit{wav2vec2-base} performs better than \textit{mandarin-wav2vec2} in the tone recognition task. For all the languages, \textit{chinese-wav2vec2-base} yields the best performance in tone recognition, with maximum F1-scores of 61.2\%, 73.9\%, and 88.03\%, achieved for Angami, Ao, and Mizo, respectively. This suggests that pre-training on a tonal language with large data enhances the model's ability to generalize tonal features to other tonal languages. However, tone recognition performance varies significantly for different languages. 
In the next subsection, we present the performance of the models in terms of their layers to understand how tone information is encoded in the SSL models. 





\begin{figure*}[h!]
\begin{center}
\includegraphics[scale=0.72]{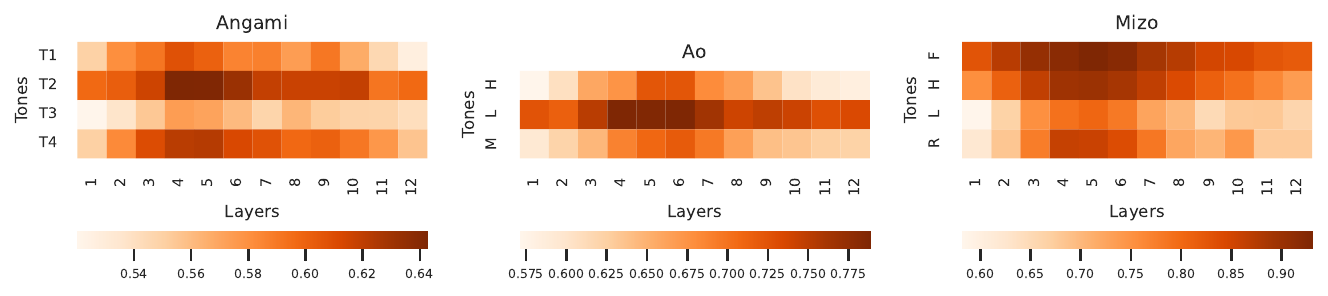}
\caption{Heatmaps illustrating tone classification accuracies (in percentage) across layers for Angami, Ao, and Mizo.}
\label{fig:heatmaps}
\end{center}
\end{figure*}

\subsection{Layer-wise performance analysis}
\label{sub:layer}
Layer-wise analysis shows that performance improves from the lower layers, peaks around the middle layers, and then declines in the higher layers for all three languages. Similar trends were also noticed in high-resourced languages, such as Mandarin in previous works ~\cite{de2024layer}.

In Angami, the performance of models is generally lower compared to Ao and Mizo. The \textit{chinese-wav2vec2-base} model still outperforms others, achieving the highest F1-score of 61.12\% at layer 4. The \textit{wav2vec2-base-vi} model follows closely, peaking at layer 7 with the F1-score of 59.80\%. The standard deviations in the F1-score provide insights into the models' consistency across different runs. It can be noticed from Figure~\ref{fig:model-wise-perf} that the F1-score has more variation in Angami for all the layers compared to Ao and Mizo.

For Ao, the \textit{wav2vec2-base-vi} model achieves the highest F1-score of 74.93\% in layer 8. The \textit{chinese-wav2vec2-base} model also shows higher accuracy, peaking at the 6th layer with an F1-score of 73.91\%. For the \textit{wav2vec2-base} and \textit{mandarin-wav2vec2}, layer number 6 and 7 yielded F1-scores of 71.94\% and 68.83\%, respectively. 

For Mizo, the \textit{chinese-wav2vec2-base} model achieves the highest F1-score of 88.03\% in the 5th layer. The \textit{wav2vec2-base-vi} model yields the highest F1-score in layer 7. Both \textit{wav2vec2-base} and \textit{mandarin-wav2vec2} models, yielded the highest F1-score in layer 4. However, \textit{wav2vec2-base} yields a higher F1-score of 80.61\% when compared to the \textit{mandarin-wav2vec2} model.

Figure~\ref{fig:heatmaps} shows the tone-wise classification accuracies (in percentage) across layers for Angami, Ao, and Mizo using heatmap plot. An interesting tone-wise patterns within each language can be seen in Figure~\ref{fig:heatmaps}. In Mizo, the \emph{F} tone generally shows higher classification accuracies across nearly all layers, appearing darker in the heatmap than other tones like \emph{L} or \emph{R}. The \emph{L} tone consistently shows lower accuracies across all layers than other tones (\emph{R}, \emph{H}, and \emph{F}), indicating that its acoustic cues are relatively harder for the model to capture. While performance on \emph{L} improves in the middle layers (4–7), it still lags behind the mid-layer peaks observed for \emph{R}, \emph{H}, and \emph{F}. The explanation for this is that the falling tone and low tone have similar $f_{0}$ contours, both exhibit a fall from a high $f_{0}$ to a lower $f_{0}$, following parallel trajectories. This indicates that the primary distinction between falling and low tones lies in their $f_{0}$ range rather than their $f_{0}$ contour. As a result, the Mizo low tone has been frequently misidentified as a falling tone 
and subsequently classified as such. By contrast, in Ao, the \emph{L} tone actually outperforms \emph{M} and \emph{H} in most layers, hinting that the model is picking up more robust or distinctive cues for the low-level pitch in Ao’s tonal system. Turning to Angami, \emph{T3} and \emph{T1} register weaker classification accuracies overall—these lighter colour blocks persist even at the mid-layers. Although these specific tone classes struggle, the middle layers (4–7) still yield the best results for all tone categories in each language.
\subsection{Dialect-independent cross-validation performance}
As discussed earlier, Angami and Ao datasets were recorded from four and three distinct dialect groups, respectively. These dialects may exhibit unique tone patterns, introducing variability and increasing the difficulty for models to accurately discriminate tones, as dialect-specific tonal characteristics may confound them. We have performed a DI CV based on tone recognition for Angami and Ao to investigate how dialectal variations affect tone recognition and to what extent the SSL-based speech representations are robust enough for this variation. In the case of Angami, the highest F1-score is 38.23\%; however, we got an F1-score of 61.12\% in the case of SI CV. It is found that a maximum average F1-score of 48.01\% is achieved for Ao, whereas, for SI CV, we obtained a maximum average F1-score of 74.93\% (Section~\ref{sub:layer}). A thorough analysis is planned in the future to better understand the effect of dialect on tone recognition for Angami and Ao.

\subsection{Discussion}
The consistent trend across languages suggests that the middle layers capture the most relevant features for tone recognition. Moreover, the drop in the F1-score in the final layers is more prominent for the non-tonal model (\textit{wav2vec2-base}) as compared to the tonal counterpart for all the languages. Also, there is a slight increase in the F1-score from layer 11 to layer 12 for all languages for \textit{wav2vec2-base}. Lower layers may focus on low-level acoustic features, while higher layers might abstract away from tonal information in favor of pre-training objectives. Researchers showed that wav2vec2 transformer layers exhibit an autoencoder-like pattern, gradually diverging from the input speech features in the initial layers and then converging back to input-like representations in deeper layers~\cite{pasad2021layer}. This layer-wise progression encodes an acoustic-linguistic hierarchy, moving from acoustic and phonetic information to lexical information (word meaning, maybe lexical tone), and then followed by a reverse trend.

The results of the current study reveal that tone inventory characteristics significantly affect the accuracy of tone recognition. The overall results indicate that tone recognition performance is highest for Mizo and lowest for Angami across all the explored models. As the pre-trained models are for Mandarin Chinese and Vietnamese, they also include information about contour tones - this may be one reason why tone recognition performance is consistently better for Mizo. Moreover, Mizo includes a mix of level and contour tones, making the tonal distinctions more pronounced than Ao and Angami, which primarily exhibit level tone systems. Additionally, Angami's four-level tone system poses a greater challenge for tone discrimination using SSL-based representations than Ao's three-tone system. Another possible explanation is related to the data conditions. Mizo is a homogenous language that has more consistent tone patterns and facilitates better model performance. However, for Angami and Ao, data were collected from three and four dialects, respectively. Different dialects may exhibit varying tone patterns, which increases the difficulty of achieving consistent tone recognition performance~\cite{Tzudir2021}.

\section{Conclusion}
The experiments demonstrate that pre-training on tonal languages and selecting appropriate layers significantly impact the performance of tone recognition systems across Angami, Ao, and Mizo. Models pre-trained on tonal languages consistently outperform non-tonal counterparts, with the middle transformer layers providing the most effective representations for capturing tonal features. The tone inventory characteristics of each language also play a significant role. Mizo, which combines level and contour tones, consistently achieves the highest recognition performance. In contrast, Angami’s more complex four-level tone system and greater dialectal variation result in lower performance, while Ao's three-tone system falls in between. These observations highlight that both linguistic factors and data variability are key determinants of system accuracy.

These insights can guide the future development of speech recognition technologies for under-resourced tonal languages, thereby enhancing linguistic inclusivity in technology. Moreover, this underscores the potential for cross-linguistic transfer of tonal representations. Given that unlabeled data for Angami, Ao, and Mizo may be relatively easy to obtain, we plan to evaluate tone recognition performance through extended pre-training of the discussed models. Furthermore, we intend to investigate how fine-tuning these models for Automatic Speech Recognition (ASR) tasks affects tone recognition performance.

\bibliographystyle{unsrt}  
\bibliography{mybib}

\end{document}